\documentclass[amsmath,amssymb, aps, prx,twocolumn,superscriptaddress]{revtex4-2}

\usepackage{braket}
\usepackage{xcolor}

\usepackage{mathrsfs,amsmath}
\usepackage{graphicx,epstopdf,epsfig}
\usepackage{bm}
\usepackage{color}
\usepackage{times}
\usepackage{hyperref}
\usepackage{physics}

\usepackage[normalem]{ulem}

\definecolor{clr}{rgb}{0,0.6,0.6}

\begin{document}
\title{Demonstration of weighted graph optimization on a Rydberg atom array using local light-shifts}
\author{A. G. de Oliveira}
\email{andre.oliveira@strath.ac.uk}
\affiliation{Department of Physics and SUPA, University of Strathclyde, Glasgow G4 0NG, UK}
\author{E. Diamond-Hitchcock}
\affiliation{Department of Physics and SUPA, University of Strathclyde, Glasgow G4 0NG, UK}
\author{D. M. Walker}
\affiliation{Department of Physics and SUPA, University of Strathclyde, Glasgow G4 0NG, UK}
\author{M. T. Wells-Pestell}
\affiliation{Department of Physics and SUPA, University of Strathclyde, Glasgow G4 0NG, UK}
\author{G. Pelegr\'i}
\affiliation{Department of Physics and SUPA, University of Strathclyde, Glasgow G4 0NG, UK}
\author{C.J. Picken}
\affiliation{M Squared Lasers Limited, 1 Kelvin Campus, West of Scotland Science Park, Glasgow, G20 0SP, UK}
\author{G.P.A. Malcolm}
\affiliation{M Squared Lasers Limited, 1 Kelvin Campus, West of Scotland Science Park, Glasgow, G20 0SP, UK}
\author{A. J. Daley}
\affiliation{Department of Physics and SUPA, University of Strathclyde, Glasgow G4 0NG, UK}
\author{J. Bass}
\affiliation{Department of Physics and SUPA, University of Strathclyde, Glasgow G4 0NG, UK}
\author{J. D. Pritchard}
\email{jonathan.pritchard@strath.ac.uk}
\affiliation{Department of Physics and SUPA, University of Strathclyde, Glasgow G4 0NG, UK}

\begin{abstract}
Neutral atom arrays have emerged as a versatile platform towards scalable quantum computation and optimization. In this paper we present demonstrations of {\color{black} solving maximum weighted independent set problems} on a Rydberg atom array using annealing with local light-shifts. We verify the ability to prepare weighted graphs in 1D and 2D arrays, including embedding a five vertex non-unit disk graph using nine physical qubits and demonstration of a simple crossing gadget. We find common annealing ramps leading to preparation of the target ground state robustly over a substantial range of different graph weightings. This work provides a route to exploring large-scale  optimization of non-planar weighted graphs relevant for solving relevant real-world problems.
\end{abstract}

\maketitle
\section{Introduction}
Optimization of combinatorially hard problems has been identified as an early potential application of quantum computing hardware \cite{abbas23}, with significant effort invested in developing protocols such as quantum annealing algorithms (QAA) \cite{farhi00,farhi01,santoro06,albash18} or variational based approaches such as Quantum Approximate Optimization Algorithms \cite{farhi14,zhou20}. Despite these efforts, hardware capable of demonstrating a practical quantum advantage in this domain remains elusive \cite{guerreschi19,lykov23,shaydulin23,bauza24}.

Neutral atom quantum computers based on arrays of individual optical tweezers \cite{saffman16,henriet20,wu21,morgado21} provide a scalable and versatile platform for quantum computing, with the ability to generate arrays of over 1000 qubits \cite{huft22,norcia24,gyger24,manetsch24} and perform high-fidelity single- \cite{nikolov23} and two-qubit \cite{levine19,evered23,ma23} gate operations, enabling realization of small scale quantum algorithms \cite{graham22}. This can be extended to implementation of logical qubit operations \cite{bluvstein24} by exploiting dynamical qubit reconfiguration \cite{bluvstein22}. Alongside digital operations, neutral atom arrays provide access to programmable spin models for performing quantum simulation \cite{browaeys20} enabling exploration of phase diagrams and dynamics in 1D \cite{bernien17} and 2D \cite{ebadi21,scholl21} and observation of novel topological phenomena \cite{leseleuc19a,semeghini21}.

\begin{figure}[t!]
  \centering
  \includegraphics[width=0.5\textwidth]{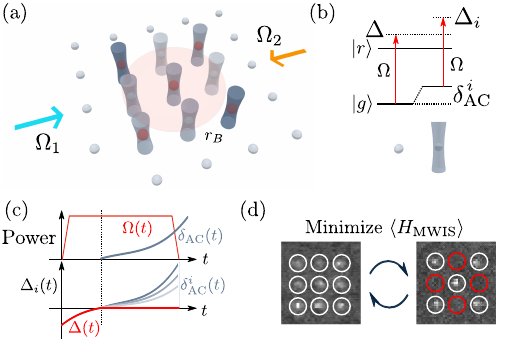} 
  \caption{Weighted graph optimization with local light-shifts. (a) The target graph problem is embedded onto a neutral atom array as a Maximum Weighted Independent Set (MWIS) on a Unit Disk Graph (UDG), using global Rydberg excitation lasers ($\Omega_{1,2}$) and an additional set of tweezer beams for implementing local light-shifts with a programmable shift on each qubit site. The Rydberg blockade mechanism prevents two atoms within distance $r_B$ being simultaneously excited. (b) Atoms without light-shift experience an effective two-photon Rabi frequency $\Omega$ and global detuning $\Delta$. The light-shift tweezers induce AC shift $\delta_\mathrm{AC}^i=w_i\delta_\mathrm{AC}$ on atom $i$ to implement a local detuning $\Delta_i=\Delta+\delta_\mathrm{AC}^i$ with relative weight $w_i$. (c) Annealing protocol using a two-stage process with the global Rydberg laser ramped from an initial negative $\Delta$ to resonance, then positive detunings defined by controlling power in light-shift tweezers to scale $\delta_\mathrm{AC}$ resulting in local detunings with fixed relative ratios defined by weightings $w_i$. (d) Ground-state solutions obtained via closed-loop optimization of the annealing profile to minimize the expectation value of the classical MWIS cost function.}
  \label{fig:setup}
\end{figure}

The ability to implement programmable spin models extends the utility of neutral atom arrays to the regime of analogue quantum computing \cite{kim23, dalyac24}, where in the context of graph optimization coupling atoms to the Rydberg states enables native embedding of the Maximum Independent Set (MIS) problem for quasi-planar 2D unit-disk graphs (UDG) \cite{pichler18,pichler18a}. A UDG is one in which only vertices within a given distance from eachtoher are connected by an edge, which can be naturally mapped to neutral atom systems using the Rydberg blockade mechanism \cite{jaksch00} which prevents more than a single Rydberg excitation occurring within a radius $r<r_\mathrm{B}$. For MIS this enforces the constraint that no two connected vertices can be simultaneously part of the same set. Experimental demonstrations of solving MIS on 2D-UDG using QAA with hundreds of qubits demonstrate the feasibility of this approach \cite{ebadi22,kim23a}, however recent analysis shows for the native Kings-graph classical solvers are able to find solutions to graphs with 1000 vertices in minutes \cite{andrist23}. One approach for extending beyond planar 2D graph is to use ancillary qubits to create quantum wires \cite{kim22} which can be routed through neighboring trapping planes in 3D to implement non-local graph connectivity \cite{dalyac23}. This approach has been used to demonstrate embedded integer factorization \cite{park24} and quadratic unconstrained binary optimization (QUBO) problems \cite{byun24} onto atom arrays using only global control.

Another approach to extend the range of problems that can be encoded using neutral atom arrays is to introduce local control fields as initially proposed in \cite{goswami24}. This permits the introduction of vertex weighting to enable solution of Maximum Weighted Independent Set (MWIS) problems on a 2D UDG, providing a route to map a variety of graph and QUBO problems onto neutral atom arrays by using a UDG-MWIS encoding. Ref.~\cite{nguyen23} provides one such approach to implementing this as a general mapping by introducing sub-graph gadgets to realize arbitrary non-local connections between vertices, with at worst a quadratic overhead in physical qubit number. A similar $\mathcal{O}(N^2)$ scaling is obtained using an alternative parity-based encoding for converted graph problems to UDG-MWIS \cite{lanthaler23}. {\color{black}{Crucially this also enables a far more efficient encoding of arbitrary problems, such as non-UDG graphs or QUBO, by mapping to UDG-MWIS with overheads $\mathcal{O}(N^2)$ than is the case for mapping to UDG-MIS with $\mathcal{O}(N^6)$ \cite{nguyen23}}}. Moving beyond the blockade regime to remove the UDG restriction offers prospects for a linear scaling in qubit overhead and application to MaxCut problems \cite{goswami24}. To date, local light-shifts have been used on neutral atom arrays in the field of quantum simulation for generating large-scale entangled states \cite{omran19}, implementing fast, arbitrary local control of interacting trimers \cite{bornet24} and studying quantum coarsening dynamics \cite{manovitz24}.

In this paper we present demonstrations of {\color{black} solving maximum weighted independent set problems} on a Rydberg atom array using annealing with local light-shifts. Our approach for weighting combines holographic techniques for generating programmable local light-shifts compatible with use on an arbitrary array geometry with the ability to implement graph annealing {\color{black}{using a dual-stage annealing protocol with}} global laser control fields, simplifying the number of available parameters when performing optimization. We verify the ability to encode weighted graphs on both 1D and 2D arrays, including embedding a 5-qubit non-UDG using 9 physical qubits. In each case we find a common annealing ramp able to prepare the target ground state solutions for arbitrary graph weightings. This work demonstrates the feasibility of implementing weighted optimization using neutral atom arrays, and provides a route to embedding non-planar graphs solving a wide range of relevant real-world problems \cite{wurtz24}.

\section{Weighted Graph Optimization}
The Maximum Weighted Independent Set (MWIS) problem on graph $G(V,E)$ comprized of $E$ edges and $V$ vertices, with each vertex having weight $w_i>0$, is defined as finding the subset of vertices offering the highest total summed weights whilst ensuring that no pair of adjacent vertices (i.e. connected by an edge) are included. Introducing the binary variables $n_i=\{0,1\}$ to indicate whether vertex $i$ belongs to the MWIS set, the classical cost function that must be minimized to solve the MWIS problem is
\begin{equation}
H_\mathrm{MWIS}=-\sum_{i\in V} w_jn_j+\sum_{(i,j)\in E} U_{ij}n_in_j,\label{eq:MWIS}
\end{equation}
where $U_{ij}>\mathrm{max}_k(w_k)$. Here the first term rewards inclusion of high-weight vertices, whilst the second term acts to penalize inclusion of vertices connected by an edge. For the case that $G$ is a UDG, this cost function can be natively mapped to the case of a neutral atom array which has the Hamiltonian given by
\begin{equation}
H_\mathrm{Ryd}=\sum_{i}(\frac{\Omega}{2}\sigma_i^x-\Delta_i \hat{n}_i)+\sum_{j>i}V(\vert \bm{r}_i-\bm{r}_j\vert)\hat{n}_i\hat{n}_j,
\end{equation}
where $\Omega$ is the Rabi frequency, $\sigma_i^x=\ket{r}_i\bra{g}+\ket{g}_i\bra{r}$, $\Delta_i$ is a site dependent detuning, $\hat{n}_i=\ket{r}_i\langle r\vert$ is the projector onto the Rydberg state of atom $i$, $\bm{r}_i$ is a position vector and $V(r)=C_6/r^6$ is the dipole-dipole interaction strength between two atoms. In the limit $\Omega=0$, the ground state of $H_\mathrm{Ryd}$ encodes the solution to the MWIS problem if $\Delta_i=w_i$ and  $V(\vert \bm{r}_i-\bm{r}_j\vert)>\mathrm{max}_k(w_k)$ for all $i,j$ sharing an edge, enabling solution of UDG-MWIS problems by annealing atoms to the ground state configuration \cite{nguyen23}.

Solving MIS and MWIS on a UDG is known to be NP-hard \cite{clark90}. State-of-the-art classical solvers for UDG-MWIS, a number of polynomial-time approximation schemes have been developed \cite{nieberg05,das20}, as well as heuristic algorithms based on iterative local-search \cite{dahlum16} or memetic algorithms \cite{grossman23} which can be applied to find approximate solutions without any guarantee of optimality. For exact solutions, MWIS can be solved using branch-and-bound methods for small to intermediate scale graphs (typically hundreds of vertices), with graph reductions enabling efficient branch-and-reduce algorithms \cite{akiba16} for scaling to larger graph instances. Reduction rules for solving MWIS on sparse graphs were introduced in \cite{lamm19} who demonstrate the ability to solve sparse real-world graphs with up to a million vertices two orders of magnitude faster than exact local-search methods, with subsequent improvements providing an algorithm running time $\mathcal{O}(1.1443^n)$ for graphs with an average degree of 3 or less \cite{huang23}. However, for the UDG-MWIS embeddings realised on Rydberg atom systems the graph is not necessarily sparse, with union-jack like connectivity giving vertices with a maximum degree of 8 meaning we typically realise a dense unit disk graph for which the reductions introduced in \cite{lamm19} for sparse graphs are expected to offer limited advantage. {\color{black}{Development of new approaches for solving UDG-MWIS using quantum hardware therefore provides an opportunity to solve large graphs without restrictions on sparsity.}}

Whilst for UDG-MIS there have been detailed studies benchmarking the scaling of state-of-the-art classical solvers against graphs with union-jack connectivity which show that classical optimal solutions can be found efficiently for graphs with thousands of vertices \cite{andrist23},  a similar analysis for UDG-MWIS remains outstanding. {\color{black}{In the results below we instead focus on experimental demonstrations of solving UDG-MWIS on neutral atom systems. We demonstrate the feasibility of the approach and outline techniques for extending this in future to tackle larger graphs in a regime competitive against classical }}algorithms.

\section{MWIS Annealing Protocol}\label{sec:MWISanneal}
Starting with a target MWIS graph problem, the graph is first mapped to a UDG-MWIS which can be geometrically embedded onto a neutral atom array using techniques from~\cite{nguyen23} with a spacing $a$ between neighbouring atoms chosen to be below the blockade radius $r_\mathrm{B}$ to ensure the blockade constraint is realized. Graph weighting is implemented with an additional light-shifting potential that is holographically projected onto the atoms using a spatial light modulator (SLM) to apply local light-shifts with arbitrary geometry and programmable relative weighting. These additional tweezer beams, shown schematically in Fig.~\ref{fig:setup}, cause a differential AC Stark shift $\delta_\mathrm{AC}^i$ on each atom proportional to the relative power in each tweezer, resulting in a total shift $\Delta_i=\Delta+\delta_\mathrm{AC}^i$, where $\Delta$ is the detuning of the global Rydberg lasers used to realize a homogeneous two-photon coupling between ground and Rydberg states. In the experiments below, the light-shift laser is blue-detuned from the atomic transitions  for both ground and Rydberg states, ensuring that $\delta_\mathrm{AC}^i\ge0$ as required. For a given graph weighting, the relative light-shifts are calibrated to realize $\delta_\mathrm{AC}^i=w_i\delta_\mathrm{AC}$, where $\delta_\mathrm{AC}$ is the global unit scale ($w=1$) of the light-shift and $w_{i}$ are the relative weights of the embedded graph problem. To avoid violating the blockade condition, it is necessary to constrain the maximum value of $\delta_\mathrm{AC}$ such that $\mathrm{max}_i(\delta_\mathrm{AC}^i)<V(a)$, where $a$ is the interatomic spacing.

To adiabatically prepare atoms in the ground state encoding the solution of the UDG-MWIS problem {\color{black} {a dual-stage annealing protocol is used, as illustrated in Fig.~\ref{fig:setup}(c). The detuning is controlled using a cubic profile (see App.~\ref{sec:AppB})}}, motivated by recent experimental demonstrations \cite{ebadi22} and features slow variation in detuning as it crosses resonance where the energy gap between ground and first excited states are minimised whilst changing faster at larger detunings where the gaps are typically larger.

Starting with the atoms initially prepared in $\ket{g}$, the global Rabi frequency $\Omega$ is ramped up using a fixed negative detuning after which the detuning is swept towards resonance. Once the global detuning reaches resonance, $\Delta=0$, the global detuning is held fixed and the remaining detuning profile is applied by dynamically varying the intensity of the light-shifting potential to control $\delta_\mathrm{AC}(t)$. By deriving the local light-shifts from a single laser, it is possible to use global intensity control with an acousto-optic modulator for implementing arbitrary detuning ramps whilst maintaining fixed relative weightings on the graph. The Rabi frequency $\Omega$ is then ramped off whilst maintaining the light-shift, after which the encoded bitstrings  $\{n_i\}$ are read out by fluorescence imaging. Atoms excited to the Rydberg state are ejected from the trap, whilst atoms in the ground state survive and appear in the images. Using a closed-loop optimization process, the cubic ramp parameters are adjusted to minimize the classical cost function $\langle H_\mathrm{MWIS}\rangle$ to maximize the probability of preparing the system in the MWIS ground state, as shown in Fig.~\ref{fig:setup}(d). Following this optimization process, the solution of the target MWIS graph problem can then be determined from the most probable output bitstrings. 

\begin{figure*}[t!]
  \centering
  \includegraphics[width=\textwidth]{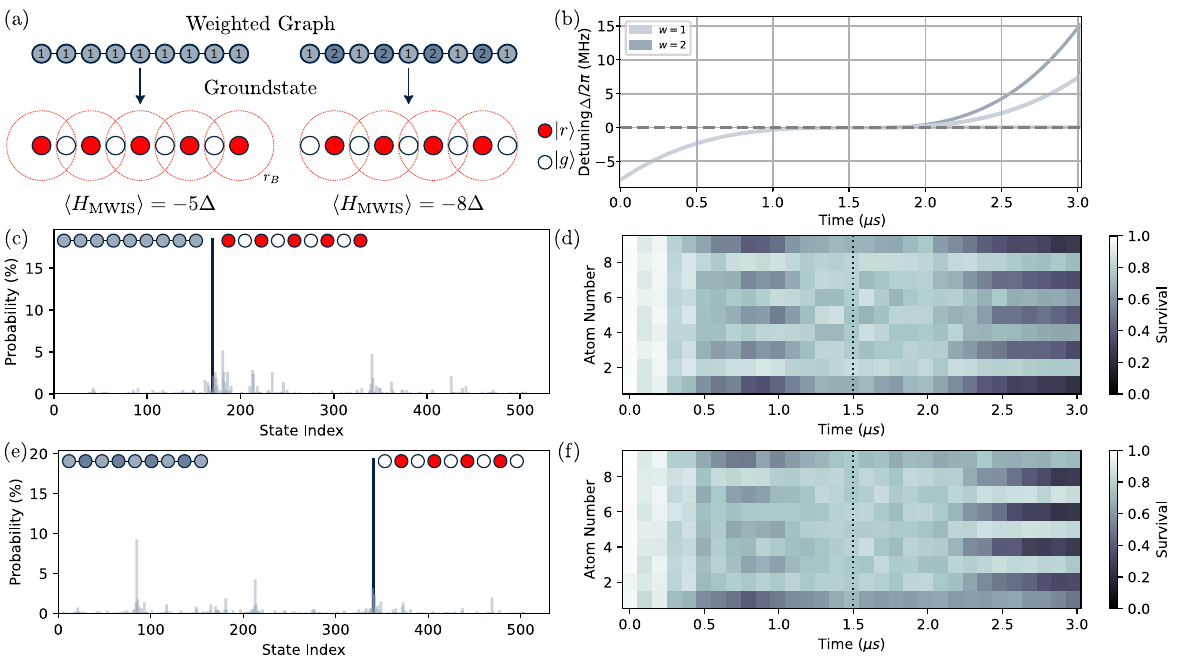} 
  \caption{1D Weighted Graphs. (a) For a uniformly weighted, odd-length 1D graph the ground state is the $Z_2$-ordered phase with Rydberg excitations on odd sites which corresponds to the unweighted MIS. Introducing a weighting with $w_i=2$ on even sites results in an MWIS ground state with Rydberg excitations localized to the even sites which is no longer equivalent to the MIS solution. (b) Optimal annealing ramp for preparing the weighted ground state obtained via closed-loop optimization for $N=9$ atoms spaced by $a=7~\mu$m. Output state probability (c) and time evolution (d) for the unweighted graph showing the odd-ordered target ground state is prepared with $19(1)\%$ probability. Output state probability (e) and time evolution (f) for the weighted graph showing even-ordered ground state is also prepared with $19(1)\%$ probability.}
  \label{fig:1D}
\end{figure*}

\section{Experimental Setup}
The experimental setup is based on arrays of holographically trapped Cs atoms. Atoms are initially prepared in the $\vert g\rangle=\vert 4,0\rangle$ state, and are coherently driven to the $\ket{r}=\ket{80s_{1/2},m_j=+1/2}$ Rydberg state by two-photon excitation via the $7p_{1/2}$ intermediate state using light at 459~nm and 1039~nm that globally illuminates the array. We operate with a detuning of $\Delta'/2\pi=$502~MHz from the intermediate state, and achieve a Rydberg Rabi frequency of $
\Omega/2\pi=2.70(2)$~MHz. Global detuning is controlled by adjusting the frequency using a double-passed acousto-optic modulator (AOM) in the 459~nm path. For $80s_{1/2}$ we calculate $C_6=-3376~\mathrm{GHz}\,\mu$m$^6$ \cite{sibalic17}, resulting in a resonant blockade radius $r_B=\sqrt[6]{\vert C_6\vert/\Omega}=10.4~\mu$m. Annealing is performed with traps off to suppress losses from the tweezers.

Local light-shifts are realized using light at 800~nm derived from a Ti:Sapph laser that is projected from a spatial light modulator (SLM) and overlapped with the main optical trapping array.  The light-shift spot arrays are imaged onto the atoms with a $1/e^2$ radius of $3.0(4)~\mu$m {\color{black}{(See App.~\ref{sec:AppA})}}. This waist size is chosen as a balance between being large enough to suppress losses from the mechanical force of the repulsive optical potential, whilst reducing cross-talk between neighbouring sites. For a given UDG-MWIS, the arrays of local light-shift beams are generated using a Gerchberg-Saxon algorithm with the desired relative weighting \cite{kim19}. To calibrate the light-shifts experienced by the atoms, we perform spectroscopy on the $\ket{50s_{1/2},m_j=+1/2}$ Rydberg state where the lower lying state suppresses the effect of atom-atom interactions. We use the measured shifts to implement closed-loop feedback on the SLM potentials to achieve target weightings with $<2\%$~RMS errors in the relative weightings after 5 iterations. The resulting differential light-shift is calibrated as $\delta_\mathrm{AC}=0.28(2)$~MHz$/$mW, which is dominated by the ground state light-shift as the Rydberg AC shift calculated using the polarizability of a free electron corresponds to only 40~kHz \cite{zhang11}.

\section{Results}
\subsection{1D Weighted Graphs}
As an artificial first demonstration of the feasibility to perform weighted optimization using the dual-stage protocol defined {\color{black}{in Sec.~\ref{sec:MWISanneal}}}, we begin with a simple and well studied problem of a 1D weighted graph as illustrated in Fig.~\ref{fig:1D}(a). For a uniformly weighted set of $N$ atoms with $w_i=1$ spaced such that $a<r_\mathrm{B}<2a$, for odd $N$ the ground state corresponds to a $Z_2$ ordered phase with Rydberg excitations localized on the odd sites as previously observed using global control only \cite{bernien17}. Introducing a relative weighting of $w_i=2$ on the even sites results in a modification of the MWIS ground state to prepare the inverted output string with Rydberg excitations now localized to the even sites.

We perform experiments on weighted 1D graphs using two 20~$\mu$m separated parallel lines of $N=9$ atoms spaced by $a=7~\mu$m, with one row set to have uniform light-shifts and the second row having $w_i=2$ on the even sites. Starting with a linear profile, we apply the global light-shift on both graphs simultaneously, and perform closed-loop feedback to minimize $\langle H_\mathrm{MWIS}\rangle$ for the weighted graph using the classical cost function defined in Eq.~\ref{eq:MWIS}. Whilst encoding the same effective ground state energy as $H_\mathrm{Ryd}$, using the classical cost function avoids issues with the optimizer trivially driving the system to large negative detunings $\Delta_i$ as a route to further minimising the energy of the observed output configurations. The resulting optimized profile is shown in Fig.~\ref{fig:1D}(b), corresponding to a duration of 3~$\mu$s with light-shifts turned on from 1.5~$\mu$s when the global Rydberg laser detuning reaches zero.

The resulting distributions of output strings are shown for both uniform and weighted graphs in Fig.~\ref{fig:1D}(c,e) respectively. Using a common annealing profile, we observe the correct ground states as the most probable state outputs in both cases, with probabilities of $19(1)\%$ in both cases. The additional 9\% peak in Fig.~\ref{fig:1D}(e) with state index 93 corresponds to the target output state with the first qubit also lost. This appears higher than other states due to slightly increased single atom loss measured for this array site. Time evolution data is shown for each graph in Fig.~\ref{fig:1D}(d,f) respectively, which clearly demonstrates the adiabatic nature of graph preparation at longer times and the emergence of the ground state ordering as expected. 

\subsection{2D Weighted Graphs}
\begin{figure*}[t!]
  \centering
  \includegraphics[width=\textwidth]{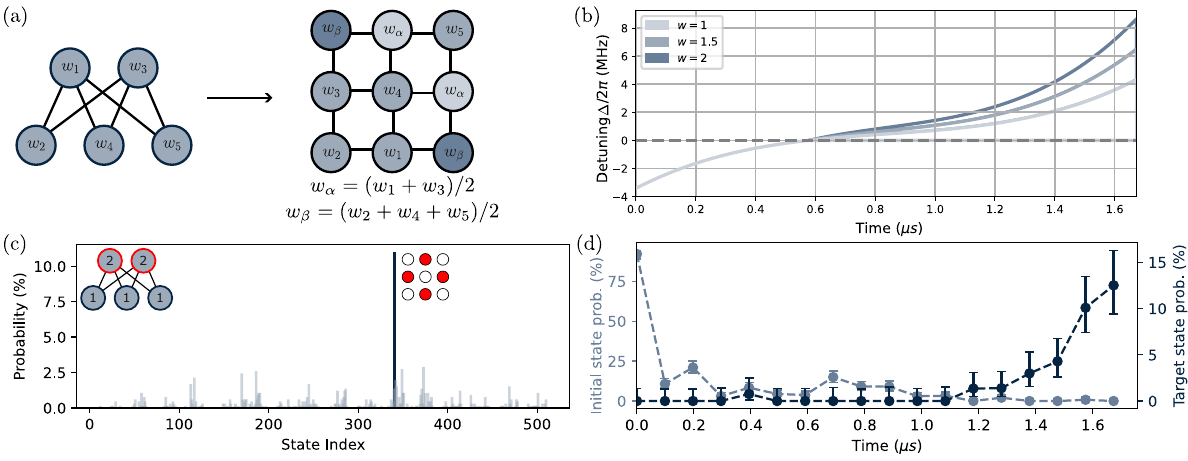}  
   \caption{2D Weighted Graphs (a) Target graph is a 5 vertex non-UDG MWIS, which can be mapped to the 9 atom UDG-MWIS graph \cite{nguyen23} by introducing four ancilla qubits with weightings $w_\alpha$ and $w_\beta$ to implement the edges from vertex 5 to vertices 1 and 3. (b) Optimized annealing profile for the graph instance with $(w_1,w_2,w_3,w_4,w_5)=(2,1,2,1,1)$ which results in $\langle H_\mathrm{MWIS}\rangle=-4.0(1)$, compared to the idealized cost of $-8$. (c) Time-evolution data for the weighted graph showing adiabatic evolution during the ramp to reach the correct target state. (d) Final output state probability distribution from 1000 experimental realizations showing the correct MWIS solution is observed with $11(1)\%$ probability, with all other output strings observed with $\le2.5\%$ probability.}
  \label{fig:2D}
\end{figure*}

Moving beyond the trivial 1D case, we consider the example of a non-UDG 2D MWIS shown  in Fig.~\ref{fig:2D}(a). This graph, featuring five vertices, has MWIS solutions corresponding to either qubits 1 and 3 being excited, or qubits 2, 4 and 5. To map this to a UDG-MWIS graph for implementation on our neutral atom array, we follow the protocol of Ref.~\cite{nguyen23} resulting in a 9-qubit graph requiring four ancilla qubits to implement the edges between vertices 1 and 2 and vertex~5. By symmetry this introduces two additional weightings, $w_\alpha$ and $w_\beta$ on the ancilla qubits. We use the weightings $w_\alpha=(w_1+w_3)/2$ and $w_\beta=(w_2+w_4+w_5)/2$ which guarantees that for any arbitrary weighting with $w_i>0$ we realize the MWIS solution as the ground state of the Rydberg interaction Hamiltonian.

Initially we consider the weighted graph instance indicated by the inset of Fig.~\ref{fig:2D}(d) with $(w_1,w_2,w_3,w_4,w_5)=(2,1,2,1,1)$ using a $3\times3$ arrangement of atoms with spacing $a=8~\mu$m. As before, we perform closed-loop optimization of the annealing profile, resulting in a shorter annealing time of only 1.65~$\mu$s with the light-shift turning on after 0.6~$\mu$s, as shown in Fig.~\ref{fig:2D}(b). For this optimization process we minimize $\langle H_\mathrm{MWIS}\rangle$ for the embedded 9 qubit graph (rather than simply just considering the non-embedded 5-qubit graph), as whilst the values of the ancilla qubits are not relevant to the problem solution, there are many possible states encoded in the 9-qubit bitstrings that feature the correct 5-vertex solution state but only one of which corresponds to the energetic ground state. This ensures the optimizer favours solutions that more closely follow adiabatic preparation of the ground states, which can be verified by observing the dynamical evolution in preparing the correct output bitstring. Fig.~\ref{fig:2D}(c) shows the preparation probability as a function of time, revealing a smooth accumulation of population after 1.2~$\mu$s and no fast oscillations observed when exploiting shorter annealing profile durations. Following optimization of ramp parameters we obtain $\langle H_\mathrm{MWIS}\rangle=-4.0(1)$, compared to an idealized cost of -8. This discrepancy in the measured cost is simply a consequence of the finite state preparation and readout fidelities leading to a distribution of output strings, with all other strings making a more positive contribution to the average cost with zero weighting contributions from atoms in the ground state and strong positive energy penalties in the case of a blockade violation.

\begin{figure}[b!]
  \centering
  \includegraphics[width=\linewidth]{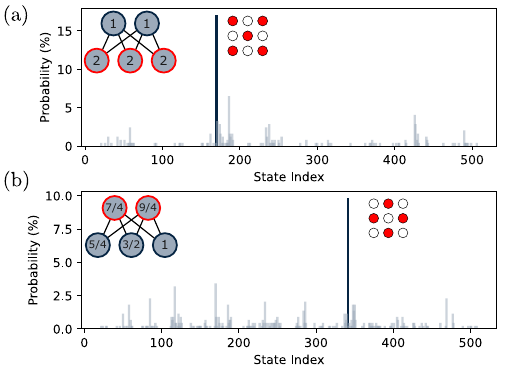}  
   \caption{Annealing alternative graph instances. We use the optimised ramp found on the $(2,1,2,1,1)$ graph and apply it to (a) $(1,2,1,2,2)$ and (b) $(7/4,5/4,9/4,3/2,1)$ graph instances. In both cases, the ramp provides the correct solution to the MWIS graph even for the case of the inverted ground state for (a), resulting in target state output probabilities of $17(2)\%$ and $10(1)\%$ with $\langle H_\mathrm{MWIS}\rangle=-6.9(2)$ and $-3.5(1)$ respectively.}
  \label{fig:2Drobust}
\end{figure}

The final output state probabilities for the weighted graph are shown in Fig.~\ref{fig:2D}(d), with the correct 9-qubit output state corresponding to Rydberg excitation of qubits mapping onto vertices 1 and 3 being observed with $\ge10\%$ probability, significantly higher than any other output bitstrings and showing demonstration of solving an embedded MWIS graph.

One further advantage of identifying an adiabatic annealing profile is that it should be robust against changes in graph weightings. To verify this for the graph shown in Fig.~\ref{fig:2D}, we directly implement the same annealing profile but with different relative weightings of $(1,2,1,2,2)$ and $(7/4,5/4,9/4,3/2,1)$. The results are shown in Fig.~\ref{fig:2Drobust} where for both cases the annealing process is able to prepare the system in the correct ground state, even for Fig.~\ref{fig:2Drobust}(a) which corresponds to the inverted solution state to that targeted previously to optimize the original annealing profile.

\subsection{Crossing Gadget}
\label{sec:gadget}
\begin{figure}[t!]
  \centering
  \includegraphics[width=\linewidth]{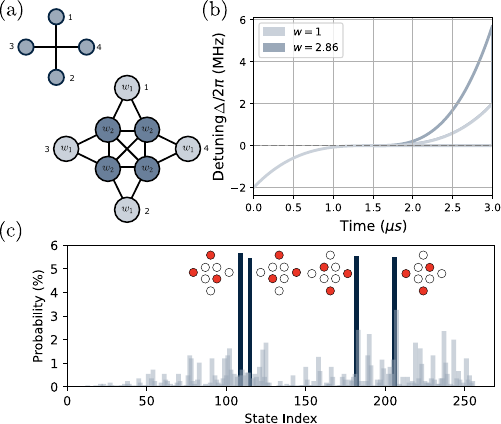}  
  \caption{ Crossing gadget. (a) Crossing gadget designed to enable logical crossing of two quantum wires connecting qubits 1 and 2, and 3 and 4, without introducing coupling between them. To compensate for interaction tails we use $w_1 = 1$ for logical qubits and $w_2=2.86$ for the ancilla atoms. The spacing between two adjacent sites with $w_2$ is $6~\mu m$, and $7.5~\mu m$  between them and their nearest $w_1$ neighbour. (b) Optimized annealing ramp (c) Final gadget output strings showing the four degenerate ground states are each prepared with probabilities of $5(1)\%$.}
  \label{fig:gadget}
\end{figure}

As discussed above, a number of proposals exist to extend the range of target optimisation problems that can be solved using neutral atom arrays by converting to a UDG-MWIS \cite{goswami24,nguyen23,lanthaler23}. One such approach introduces simple \textit{gadgets} that provide a generalised framework for mapping problems to UDG-MWIS  \cite{nguyen23}, and here we consider a proof-of-concept implementation of the crossing gadget using our local light-shift approach. This crossing gadget, shown in Fig.~\ref{fig:gadget}(a), implements an edge between qubits 1 and 2, and qubits 3 and 4. This requires four additional atoms, where the logical qubits (1-4) are assigned weight $w_1=1$, whilst the ancilla gadgets have $w_2=2.86$ to ensure the lowest energy state corresponds to an equal superposition of the four degenerate output configurations with either qubit 1 or 2 excited independent of 3 or 4. This deviates from $w_2=4$ presented in \cite{nguyen23} due to the requirement to compensate for the effects of the $1/R^6$ interaction tails on the ancilla atoms {\color{black}{(See App.~\ref{sec:AppC})}}. For these measurements the intermediate state detuning was increased to $\Delta'/2\pi=1$~GHz and Rabi frequency reduced to $\Omega/2\pi=1.28$~MHz motivated by numerical modelling of the annealing process. As above, we then optimize the annealing ramp experimentally to minimize $\langle H_\mathrm{MWIS}\rangle$ for the problem resulting in the ramp shown in Fig.~\ref{fig:gadget}(b). The distribution of the final output strings are shown in Fig.~\ref{fig:gadget}(c), where the four target ground states are the most likely to be observed with approximately equal probability, and $\ge 20\%$ combined probability. Analysis of the next four most probable output strings show these correspond to the case of a target ground state with one of the logical qubit Rydberg excitation's missing, likely due to limitations in the Rydberg detection fidelity we estimate a detection error of 8.5\% for the current detection method based on ejection from the tweezer potentials \cite{leseleuc18}). For embedding of larger graphs, improvements in the Rydberg state detection can be implemented using microwave ionization \cite{graham19,ebadi21}, however these results verify the proof-of-concept approach for performing gadget encodings on the current architecture.

\section{Discussion and Outlook}
In this paper we have demonstrated a new approach to implementing UDG-MWIS using arbitrary local light-shifts implemented using a SLM combined with a dual stage annealing profile that combines the advantages of local light-shifts with global frequency and intensity control for implementing a quantum annealing algorithm. We demonstrate the ability to encode ground states of small-scale 1D and 2D graphs, including demonstrating the embedding of a non-UDG MWIS problem. In both cases, we were able to identify robust annealing protocols suitable for use with different relative weightings providing the possibility to design universal ramps that are suitable for solving a large number of graphs without the required overhead of full closed-loop optimization of the annealing ramp each time.

This approach for local light-shifts is highly scalable, with SLMs able to generate arrays of $>\!\!1000$ individually controllable spots and the option to implement in-situ realtime feedback using a camera calibrated against the measured atomic shift previously demonstrated for normalization of arbitrary trapping potentials \cite{bluvstein24}. Currently at 800~nm we have up to 4~W available at the experiment, providing sufficient power for $>\!\!100$ weighted sites. In future, this can be scaled further using coherent combination or frequency doubling telecoms fiber lasers to increase the available power and combined with techniques for beam-shaping to implement a flat-top point spread function on the SLM \cite{schroff23} to reduce the size of the light-shift spots and further reduce mechanical effects caused by using a repulsive optical potential to enable longer annealing runs. For larger graphs the system also requires increased uniformity of Rydberg excitation over the array which can be implemented using additional SLMs in the global Rydberg beam paths to create top-hat intensity profiles \cite{ebadi21}.

The measured output state probabilities in the current experiments are limited to 10-20\% by a number of factors. Fundamental limits arise from the finite-duration of the annealing profile and spontaneous decay from the Rydberg state alongside scattering from the intermediate excited state, with numerical simulations predicting target output probabilities between 50-60\% for the graphs used here. This is further reduced by the effects of finite state preparation fidelity, finite Rydberg detection fidelity and additional losses such as finite vacuum lifetime. By implementing enhanced optical pumping techniques \cite{levine19}, microwave-assisted Rydberg detection \cite{graham19,ebadi21}, extended vacuum lifeitmes \cite{manetsch24} and increased laser power to suppress losses by facilitating larger intermediate state detunings \cite{evered23} these limitations can be suppressed to maintain high-fidelity readout for large arrays.

For larger graph problems, the quadratic overhead introduced by the UDG-MWIS encoding \cite{nguyen23} introduces interesting open questions around scalability of this approach, whilst recent theoretical work exploring annealing of unweighted graph shows even for relatively simple graph constructions it is possible to have minimum energy gaps that decay superexponentially with system size \cite{schiffer23} requiring use of optimal control techniques or quenches to prepare atoms in the target ground states. In future work we will explore this question further, and provide a comparison against other encoding protocols requiring local light-shifts as a route to identifying useful applications of weighted graph optimization on neutral atom arrays \cite{lanthaler23,goswami24}.

\begin{acknowledgements}
We thank H. Pichler for helpful discussions. This work is supported by the EPSRC Prosperity Partnership \emph{SQuAre} (Grant No. EP/T005386/1) with funding from M Squared Lasers Ltd and EPSRC Grant EP/Y005058/2. The data presented in this work are available at \cite{oliveira24data}. 
\end{acknowledgements}

\appendix

\begin{figure}[b!]
  \centering
  \includegraphics[width=\linewidth]{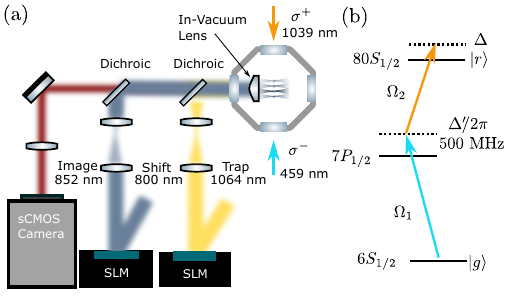}  
   \caption{Experimental Setup. (a) Atom arrays are generated using 1064~nm trapping light projected into the chamber using an SLM and a high-numerical aperture in-vacuum aspheric lens. Local light-shifts are implemented using a second SLM to overlap spots at 800~nm with the 1064~nm tweezers. Atom imaging is performed by collecting fluoresence onto an sCMOS camera. Rydberg excitation is performed using counter-propagating beams at 459~nm and 1039~nm with orthogonal circular polarisiations which globally illuminate the array. (b) Two-photon Rydberg excitation scheme from the $6S_{1/2}$ clock state $\vert g \rangle=\vert 4,0\rangle$ to $\vert r\rangle=\vert ns_{1/2},m_j=+1/2\rangle$ via the intermediate $7P_{1/2}$ level. Lasers are locked with a detuning of $\Delta'/2\pi=506$~MHz from the $7P_{1/2}$ transition.}
  \label{fig:sup1}
\end{figure}

\section{Experiment Setup}\label{sec:AppA}
\emph{Neutral Atom Arrays}. --- A schematic of the experimental setup is shown in Fig.~\ref{fig:sup1}(a), based on the setup previously introduced in Ref.~\cite{nikolov23}. Atom arrays are created by loading individual Cs atoms into 1064~nm optical tweezers generated using a spatial light modulator (SLM). Traps are focused to a $1/e^2$ waist of 1.5~$\mu$m using an in-vacuum aspheric lens with NA=0.45 \cite{sortais07} which has an ITO coating to suppress charge build up. This same lens is used for qubit readout by performing fluorescence imaging with an sCMOS (Teledyne Photometrics Prime BSI) camera.

\begin{figure*}[!htbp]
  \centering
  \includegraphics[width=\textwidth]{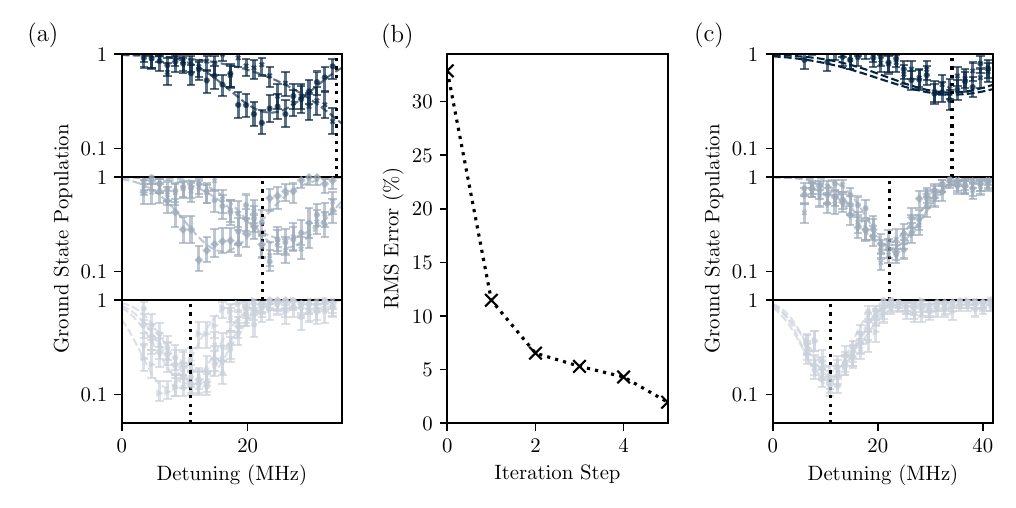}
  \caption{Feedback on local light-shift potentials. (a) Spectroscopy of initial hologram generation for the $3x3$ atom arrangement with $(w_1,w_2,w_3,w_4,w_5)=(1,2,1,2,2)$ requiring $w_{\alpha}=2$ and $w_{\beta}=3$ shows significant variation in the shift experienced by each atom. (b) Performing closed-loop feedback on the SLM hologram generation using the measured light shifts enables rapid convergence onto the target light shifts with RMS below $<2\%$. (c) Spectroscopy measurement performed after iteration 5 showing close convergence of the relative shifts experienced by each atom. The dashed line shows the targeted resonance for each atom.}
  \label{fig:fbk}
\end{figure*}

Following stochastic loading of the array, we rearrange the atoms into the target spatial configuration with a mobile tweezer beam at 1039~nm created using a dual-axis acousto-optic deflector (AOD) driven by a fast micro-controller. Atom sorting is performed using a trap depth of 1~mK and a tweezer of 9.8~mK, using a customised algorithm to calculate moves in real time that aims to minimise the number of atoms movements required to fill the target array sites. We achieve single move efficiencies $\ge97\%$, and typical success rates $\ge70\%$ for the graphs shown above using a single round of sorting. For larger graphs, multiple sorting iterations can be performed.

After rearrangement, atoms are optically pumped into the $\ket{4,0}$ hyperfine state using $\pi$-polarized light on the Cs $D_1$ transition resonant with the $F=4\rightarrow F'=4$, achieving an optical pumping efficiency of $0.971(9)$ measured using microwave benchmarking \cite{nikolov23}.

For the experiments in this paper, prior to Rydberg operations the optical dipole traps are ramped down to 15~$\mu$K, cooling the atoms to temperatures of 2~$\mu$K. Atoms are then released from the traps for 5~$\mu$s to minimise the mechanical force of the tweezer beams on the Rydberg states, and atoms are then recaptured at 1~mK to facilitate Rydberg state detection via ejection of the Rydberg states from the traps. For the current parameters we estimate a Rydberg detection error of 8.5\% \cite{leseleuc18}. 

\emph{Rydberg Excitation}. ---
Rydberg excitation is performed using two-photon excitation via the intermediate excited state as shown in Fig.~\ref{fig:sup1}(b) using light at 459~nm and 1039~nm to drive transitions from $6P_{1/2}\rightarrow7P_{1/2}$ and $7P_{1/2}\rightarrow nS_{1/2}$ respectively. Both beams are derived from Ti:Sapph lasers (M Squared SolsTiS) at 918~nm and 1039~nm which are frequency stabilised to a commercial ULE cavity (M Squared) offering finesse $\mathcal{F}>40$~k at both wavelengths. The lasers are offset-locked using a Pound-Drever-Hall with fiber EOMs \cite{legaie18} to provide controllable detuning with respect to the 3~GHz spaced cavity modes, with few kHz linewidths measured via a beatnote between the two lasers.

The 459~nm light is frequency doubled (M Squared ECDF) and the detuning controlled using a double pass accousto-optic modulator (AOM,AA MQ180-A0.25-VIS) in a cat-eye geometry to enable fast frequency control. The AOM is driven using a high bandwidth arbitrary waveform generator (AWG, Spectrum M4i6631-x8). This is delivered to the experiment by fiber, and overlapped onto the array with a 125~$\mu$m $1/e^2$ waist. The 1039~nm laser seeds a fiber amplifier (Azur ALS-1039-20-CP) which is intensity controlled using a single-pass AOM and coupled into a high-power fiber. The beam is then imaged onto the atoms using cylindrical optics to create a beam with a $1/e^2$ waist size of $60\times20~\mu$m. For the experiments presented here, Rydberg excitation is performed using an intermediate detuning of $\Delta'/2\pi=502$~MHz with respect to the $7P_{1/2}$ centre of mass energy, with powers set to provide a Rabi frequency of $\Omega/2\pi=2.70(0)$~MHz over the central rows of the array.

To stabilise the intensity, prior to each experimental cycle a 1~ms sample pulse is applied allowing closed-loop feedback to an AOM based noise eater on each laser \cite{preuschoff20}, with the intensity held constant for the fast Rydberg pulses. We achieve $1\%$ shot-to-shot stability stable over hours using this technique. 

\emph{Local Light-shifts}. --- To apply local light-shifts for implementing graph weighting, an 800~nm laser (M Squared SolsTiS) is used with a second spatial light modulator to create a secondary tweezer array that is overlapped with the underlying 1064~nm traps as shown in Fig.~\ref{fig:sup1}(a). Due to the chromatic shifts of the in-vacuum aspheric, it is not possible to image both 1064~nm and 800~nm arrays overlapped after the chamber focused in a common plane. Instead, to overlap the arrays we exploit the fact that ground-state atoms are anti-trapped by this blue detuned light. Using this effect, we initially generate arrays with a 2~$\mu$m waist and apply the 800~nm potential onto the trapped atoms for 1-10~ms, resulting in ejection of atoms from sites that are overlapped between the two beams. Using $10\times10$ arrays with 8~$\mu$m spacing, we perform multiple iterations to adjust relative spacing, rotation and defocus of the 800 spots with respect to the 1064~nm array using the SLM to maximise the relative alignment of the two arrays by minimising atom survival. This also ensures the 800~nm light is focused on the same plane as the 1064~nm traps.  

For performing local-light shifts, we suppress the mechanical effects of the repulsive light shift on the ground-state atoms by changing the input beam size on the SLM to create spots with an effective waist size of 3.0(4)~$\mu$m, resulting in $<2\%$ additional atom loss when applying local shifts of up to 10~MHz for periods of up to 3~$\mu$s. In the present work this is achieved by masking the active area of the SLM to reduce the effective beam size at the cost of reduced efficiency and limitations in the number of spots that can be generated due to finite SLM pixels. In future this limitation can be readily overcome by adapting the magnification of the relay optics from SLM to the atoms enable high-efficiency generation of arrays of over 1000 spots \cite{kim19}.

Light-shift potentials are generated using the Gerchberg-Saxton algorithm \cite{kim19}, with the modification that the relative weighting of each spot is adjusted to reflect the target weighting $w_i$ of the specific graph problem. Following the initial hologram generation, we perform Rydberg spectroscopy at $n=50$ to minimise shifts due to interactions to measure the resulting $\delta_\mathrm{AC}^i$ for each site. An example spectroscopy measurement is shown in Fig.~\ref{fig:fbk}(a) for the case of the $(w_1,w_2,w_3,w_4,w_5)=(1,2,1,2,2)$ graph requiring $w_{\alpha}=2$ and $w_{\beta}=3$. This data shows that whilst sites initially defined with higher weightings have larger shifts, there is a large RMS error between the initial hologram generation and the target light shifts. To rectify this, we implement closed-loop feedback on the relative weightings using the same approach used for normalising relative intensity in trap arrays \cite{kim19,ebadi21}. Following each spectroscopy measurement, an additional hologram generation step is taken using updated weights to drive convergence towards the target light-shifts. Fig.~\ref{fig:fbk}(b) shows this can be used to obtain RMS errors $<2\%$ after 5 iterations, with the measured light-shifts after the final iteration shown in Fig.~\ref{fig:fbk}(c) highlighting the ability to obtain accurate light shift potentials.

Note that this same SLM based approach could be used to create local light shifts using red-detuned light, with the requirement that now vertices with low weight require higher power light shifts and global detuning and light-shift beams need to be modulated simultaneously during the positive detuning part of the ramp. Compared to our implementation with blue-detuned light, this approach both places greater demands on the bandwidth of the AOM driving the global detuning and also couples the effects of a finite AOM bandwidth and finite local light-shift beam power in determining the maximum achievable total detuning. The use of red-detuned light shift beams is thus more technically challenging and, while mitigating the anti-trapping effect of the light shift beams on the atoms in their ground state, still has an anti-trapping effect on the Rydberg state atoms

\section{Weighted-graph Annealing}\label{sec:AppB}
Optimization of weighted graphs is performed using a dual-stage annealing profile as illustrate in Fig.~\ref{fig:setup}(c). To simplify optimization of the annealing profile, we parameterize the detuning using a cubic ramp of the form \cite{bernien17}
\begin{equation}
\delta(t)=at^3+bt+c,
\end{equation}
with ramp parameters $a=8s c/\tau^3$, $b=2c/\tau-a\tau^2/4$ and $c=(\Delta_\mathrm{max}-\Delta_\mathrm{min})/2$. Here $0\le s\le1$ acts as a shape parameter, with $s=0$ equivalent to a linear ramp, and $\tau$ is the total ramp duration.

This profile is then used to build piece-wise profiles, with the global Rydberg laser detuning $\Delta(t)$ and 800~nm light-shift $\delta_\mathrm{AC}$ given by
\begin{equation}
\Delta(t) = 
\left\{
    \begin{array}{ll}
        \delta(t) & \text{if } \delta(t)<0,\\
        0 & \text{otherwise,}
    \end{array}\right.
    ,\delta_\mathrm{AC}(t) = 
\left\{
    \begin{array}{ll}
        0 & \text{if } \delta(t)<0,\\
        \delta(t) & \text{otherwise.}
    \end{array}\right.
\end{equation}

To synchronise the pulses, the parameters are sent to the same dual channel AWG with one channel controlling amplitude and frequency of the 459~nm laser, and the second channel connected to the 800~nm AOM situated before the SLM that provides fast intensity control.

For the graphs shown above, annealing profile parameters $s$, $\tau$, $\Delta_\mathrm{min}$ and $\Delta_\mathrm{max}$ are optimized using MLOOP \cite{wigley16}, an open source machine learning package interfaced through Python. For each set of parameters, we perform 200-300 repeated measurements to evaluate the averaged classical cost function $\langle H_\mathrm{MWIS}\rangle$ from the distribution of experimentally observed bitstrings, and use this cost to control the evolution of the optimization process. Following optimization, we perform 1000 shot repeats to verify the final output probabilities presented in the paper.

\begin{figure}[b!]
  \centering
  \includegraphics[width=\linewidth]{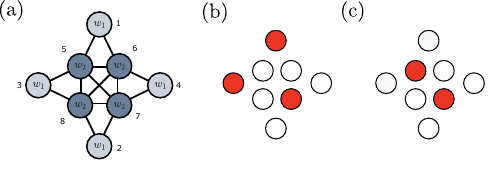}  
   \caption{Crossing gadget weightings. (a) Crossing gadget schematic showing labels for weight and qubit number. (b) One of the four possible target output states that encodes the UDG solution (c) Actual ground-state for the case of uniform spacings (i.e. $r_{15}=r_{56}$), $w_1=1$ and $w_2=4$ when including real $1/R^6$ interaction tails.}
  \label{fig:sup3}
\end{figure}

\section{Crossing Gadget Weightings}\label{sec:AppC}
In Sec.~\ref{sec:gadget} above we demonstrate use of the crossing gadget initially introduced by \cite{nguyen23} and shown schematically in Fig.~\ref{fig:sup3}(a) with qubit labels. In the UDG limit, where we consider only the interactions between qubits connected by edges, the target ground-state can be encoded using weights $w_1=1$ and $w_2=4$ resulting in a cost of $C=-2w_1-w_2=-6$ for each of the four degenerate output states as demonstrated in Fig.~\ref{fig:sup3}(b).

When embedded on the realistic neutral atom system however with interactions of the form $V(R)=-C_6/R^6$, the energy of this target configuration becomes $E_0 = -2w_1\delta_0-w_2\delta_0+V(r_{13})+V(r_{17})+V(r_{37})$. In the case that all edges are equal distance, $w_1=1$ and $w_2=4$, then the true groundstate of the system becomes that shown in Fig.~\ref{fig:sup3}(c) with energy $E'=-2w_2\delta_0+V(r_{57})$.

To mitigate this issue and ensure $E_0 < E'$, we adjust the spacings to make $r_{75}\simeq r_{15}$ and choose $w_2$ to maximise the energy gap between $E_0$ resulting in choice of $w_2=2.86$ and $r_{75}=0.8r_{15}$ in the experiments above.

\end{document}